# Odd-even effect in *n*-alkane systems: A molecular dynamics study

I. Dhiman[a], Marcella C. Berg[b,c], Loukas Petridis[c], Jeremy C Smith[c], S. Gautam[d]

[a]*Centre for Energy Research, Budapest, 1121 Hungary*

[b]*Jülich Centre for Neutron Science (JCNS) at Heinz Maier-Leibnitz Zentrum (MLZ), Forschungszentrum Jülich GmbH, Garching, Germany*

[c]*Oak Ridge National Laboratory, Center for Molecular Biophysics, Oak Ridge, TN, 37831, USA*

[d]*School of Earth Sciences, The Ohio State University, 275 Mendenhall Laboratory, 125 S Oval Drive, Columbus, OH, USA*

**Abstract**

Alternation in various properties of *n*-alkanes ($C_nH_{2n+2}$) as a function of carbon content (*n*) is termed 'odd-even effect'. Here, we report a comprehensive molecular dynamics simulation study on *n*-alkane systems carried out with *n* ranging between 3 (propane) and 8 (octane), examining the odd-even effect in melting point, density, intramolecular conformational ordering, translational and rotational motion. We observe an odd-even alternation in these properties, but with heptane (*n* = 7) exhibiting anomalous behavior for all except conformational ordering. Our simulations also show the presence of odd-even behavior in rotational and translational dynamics, below and above the melting point, respectively. The results highlight the role of both molecular shape and the variation in density and their interplay in the origins of the odd-even effect.

*Email addresses:* id.indudhiman@gmail.com (I. Dhiman), ma.berg@fz-juelich.de (Marcella C. Berg), gautam.25@osu.edu (S. Gautam)

# 1.0 Introduction

Several properties of long chain hydrocarbons show an alternating variation with the number of carbon atoms, which is termed the 'odd-even' effect [1-6]. In general, odd-even alteration behavior is correlated with the difference in packing density for long ranged periodically packed molecules, wherein the odd numbered alkanes ($C_2$-symmetry) do not pack as densely as the even numbered ones ($C_i$-symmetry) [5]. This effect is correlated with the physical properties of various homologous systems, particularly in *n*-alkanes ($C_nH_{2n+2}$) and their $\alpha$- and $\alpha, \omega$- substituents [7 - 10]. Also, the effect is more pronounced for shorter chains, with the longer the chain length, the smaller the difference in melting points between consecutive odd and even numbered *n*-alkanes.

The understanding of odd - even dynamics in alkanes is useful from an industrial application viewpoint. Particularly, in the petroleum industry the transport behavior of alkanes is crucial to understand the chemical processes involved in recovery and refinement [11-15].

Extensive studies have been reported aimed at understanding odd-even behavior in properties, such as the density, modulus, viscosity, sublimation enthalpy and mechanical and surface properties [1-5, 9, 10]. The earliest work in this field was carried out by Baeyer et al. in 1877, wherein odd-even based alteration of melting point as a function of chain length for fatty acids was discovered [1]. In 1999 Boese et al. reported this behavior of melting point alteration in short chain *n*-alkane systems [5]. Both Boese et al. [5] and Thalladi et al. [9, 10] proposed a geometry based - parallelogram (for even numbered alkanes) - trapeziod (for odd numbered alkanes) model to explain this effect. Thereafter several studies were conducted on the odd-even behavior in chain hydrocarbons [9, 10, 16-18], such as on ionic glasses [20] and liquid *n*-alkane systems etc. [21]. Similar behavior was also observed in their derivatives such as $\alpha, \omega$- alkane-diols, diamides, diamines and dicarboxylic acids [2, 9, 10, 22-26].

Although the odd-even effect is well known and observed in various systems, there is still a lack of understanding of the dynamical basis of odd-even alteration. Many of the experimental studies indicate the requirement for a detailed understanding of the odd-even behavior at molecular level for which molecular dynamics (MD) simulations are well suited [27-36]. An MD simulation study performed on a 4-*n*-alkyl-4'-cyanobiphenyl liquid crystal showed a slight odd-even effect in the rotational diffusion coefficient, rotational viscosity coefficient and molecular ordering parameter [27]. Also, an MD study performed on ionic liquid systems is consistent with the behavior observed for the diffusion coefficient and electrical conductivity in ionic liquid systems [30, 31]. Another recently reported combined NMR and computational study highlights the importance of dynamic behavior in homologous systems [18]. In the literature, several studies have been reported investigating the correlation between structure and physical properties, such as surface tension, volatility and dissociation energies [6, 32-36].

In this work, our aim is to investigate the odd-even effect on dynamical behavior of molecules in *n*-alkane systems and further to elucidate the physical properties that drive this effect. We have performed MD simulations on *n*-alkane systems for *n* = 3 - 8. To understand the dynamical properties of n-alkanes around their melting points, we performed simulations at several temperatures with ± 5 K steps above and below the melting points. Initial estimates of the melting point of each alkane were derived from the experimental work reported in literature. To understand the dynamical properties of *n*-alkanes around their melting points, we performed simulations at several temperatures with ± 5 K steps above and below the melting points. Initial estimates of the melting point of each alkane were derived from the experimental work reported in literature. With addition of every $CH_2$ group in *n*-alkane series the intermolecular distance between the end $CH_3$ groups is increased [37]. This in turn influences the molecular structure in the liquid state. Probing several different quantities related to the structure and dynamics pertaining to both orientational as well as positional order, helps us point to those properties that are the most dominant contributors to the origin of the effect.

## 2.0 Simulation Details

We have carried out classical MD simulations using the DL-POLY classic [39, 40] on a series of *n*-alkanes (*n* = 3 - 8), using the TraPPE-UA convention [38]. The *n*-alkanes (*n* = 3 - 8) selected are listed in Table 1. A united-atom and semi - flexible model was used for all the *n*-alkane molecules [38], where all the $CH_3$ and $CH_2$ groups in a molecule are treated as a single unit. The TraPPE-UA force field was used, which provides a method for simulating alkanes in a united atom approach. The reason for this is twofold [38]. Firstly, ambiguity related to the hydrogen flexibility in the resulting data can be avoided. Secondly, this also has the advantage of being computationally less expensive in comparison to an all-atom simulation. The initial crystallographic structural information for all the *n*-alkane molecules was obtained from the Cambridge crystallographic data centre (CCDC) and by the x-ray diffraction work carried out by Boese et al. [5] Each individual simulation cell was built using the VESTA software [41]. The whole system was repeated in such a way that final size of the cell is at least 28 Å along all the directions of the unit cell. Preliminary runs for all the temperatures in the NPT ensemble for 1.2 ns were carried out to estimate the density at the given temperature. Thereafter, these density values were used to carry out the production simulations in the NVT ensemble. The system equilibration was confirmed by ascertaining that the fluctuations in temperature and pressure are below 5 %, and with energies (potential and kinetic) stabilized.

All the NVT simulations were performed in two parts to capture processes occurring at both short (sub picosecond) and long (nanosecond) timescales. At shorter timescale, the simulations were performed for 1.2 ns with trajectories saved for every 0.010 ps, while at longer timescale the simulations were carried out for 10 ns with trajectories written at every 0.5 ps. The initial run time of 0.5 ns and 1 ns for the respective short and long runs is considered as equilibration time. For unwrapping the trajectories the VMD software tool was utilized [42]. Long-range interactions were calculated by a 3D Ewald sum. Three-dimensional periodic boundary conditions were used and following the TraPPE-UA convention 14 Å is taken as cut-off. All simulations employed an integration step of 1 fs. Further simulation details and unit cell dimensions and the related multiplicity values are described in the supplemental and Table S1.

**Table 1:** The properties of *n*-alkane molecules from *n* = 3 - 8 comparing experimental and simulated values are summarized. The experimental melting point values listed in the table are based on the [5, 37].

| Molecule | Formula ($C_nH_{2n+2}$) | Molecular schematic | Experimental melting point (K) | Simulation melting point (K) | Experimental density* g/cc | Simulation density g/cc |
|---|---|---|---|---|---|---|
| Propane | $C_3H_8$ | 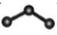 | 85 | 100 | 0.73199 | 0.726 (1) |
| Butane | $C_4H_{10}$ | 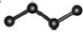 | 135 | 140 | 0.73407 | 0.737 (2) |
| Pentane | $C_5H_{12}$ | 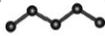 | 143 | 125 | 0.7624 | 0.784 (2) |
| Hexane | $C_6H_{14}$ | 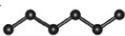 | 178 | 165 | 0.7618 | 0.778 (1) |
| Heptane | $C_7H_{16}$ | 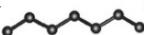 | 182 | 195 | 0.77585 | 0.778 (2) |
| Octane | $C_8H_{18}$ | 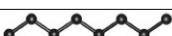 | 216 | 200 | 0.76384 | 0.788 (2) |

*Experimental density values are taken from the following link https://webbook.nist.gov/chemistry/fluid/, while the simulated densities are obtained from NPT runs performed here.

## 3.0 Results

### 3.1. Evolution of structure with temperature

The odd-even behavior is depicted in thermodynamic properties, such as melting point and density, of *n*-alkanes with increasing chain length (*n*). To obtain the melting point of each system from the simulations at each temperature, we calculated the orientational order parameter as a function of time using equation 1,

$$s(t) = \langle \frac{3}{2} cos\theta(t) - \frac{1}{2} \rangle \qquad (1)$$

where, $\theta(t)$ is the angle made by the molecular axis with the reference direction shown in the inset to Figure 1(a) and $\langle \ \rangle$ denotes an ensemble average. The temperature at which the order parameter values change from -0.5 or 0.2, i. e. from anti-parallel or partial alignment with respect to the reference direction, to 0, i. e. with completely random alignment, was denoted as the melting point. As an example the calculated order parameter for pentane for different temperature is shown in Figure 1(a). The calculated orientational order parameter as a function of time at various temperatures for all the alkanes is shown in the supplemental in Figure S1.

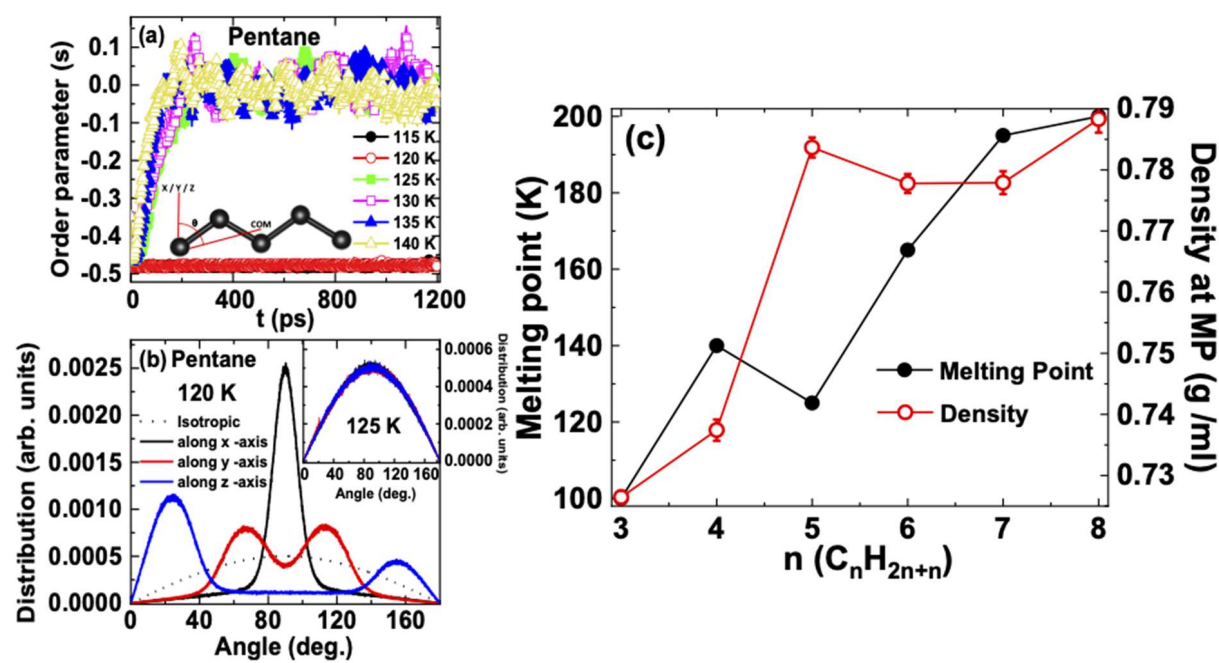

Figure 1: (a) Calculated orientational order parameter (s) as a function of temperature for pentane. The molecular axis with respect to the reference direction and Cartesian coordinates is shown in the inset. (b) Representative temperature dependence of orientational distribution function along X (in black), Y (in red) and Z (in blue) axis for pentane below (120 K) melting point and above melting point (125 K) in the inset. (c) Melting point (in black) and density (in red) of $n$ - alkanes for $n$ = 3 - 8. The density values are obtained from simulations performed in NPT ensemble, whereas the melting point is calculated from the order parameter in a NVT ensemble. The odd - even behavior in both melting point and density is clearly evidenced, with heptane ($n$ = 7) showing anomalous behavior. Lines serve as guide to the eye.

The orientational structure was also probed by calculating the distributions of molecules with their molecular axis aligned at an angle ($\theta$) with respect to the Cartesian directions, as depicted in the inset to Figure 1(a). This distribution, henceforth referred to as the orientational distribution function (ODF), along with the isotropic distribution expected in the absence of orientational order is shown in the supplemental section in Figure S2 for all the molecules. The transition from an ordered state represented by sharp peaks in the ODF (such as for pentane at 120 K in 1(b)), to a disordered state marked by an ODF that closely follows the isotropic distribution, i.e. forms a perfectly random distribution can be seen. This transition also marks the melting point (e.g., inset to 1(b) at 125 K).

The density values were extracted from the simulations performed in NPT ensemble, as described in the simulation details section. In Figure 1(c) the melting point and density as a function of $n$ are shown. For lower $n$-alkanes ($n$ = 3 - 6) clear odd – even behavior in melting point and density is evidenced, while at higher values of $n$ = 7 and 8, this effect is less prominent. Weakening of odd - even behavior with increasing value of $n$ has been reported for other systems, like diamides [9]. However, the deviation for $n$ = 7 from the expected odd-even behavior, as seen in Figure 1(c), may also be influenced by to the parameterization of force field for $n$ = 3, 4, 5 and 8 alkanes, as explained in the discussion section.

3.1.1. Positional order

To further examine structural information as a function of temperature, radial distribution functions (RDF) for $CH_3$ - $CH_3$ pairs belonging to different molecules are plotted in Figure 2. The temperature dependence of the RDF for $CH_3$ - $CH_2$ and $CH_2$-$CH_2$ pairs for all the molecules are shown in the supplemental Figures S3 and S4, respectively. For all the alkanes, the transition from an ordered state represented by the presence of sharp peaks in RDF to a disordered state shown by shallower and broader peaks can be seen. In the ordered state, i. e. below their respective melting points for all the molecules, the slight broadening observed in the sharp RDF peaks can be attributed to the presence of rotational disorder, as well as angular and dihedral vibrations in the system. Increasing the temperature displays a trend towards the disordered state.

In Figure 2(c) the temperature dependent RDF for pentane $CH_3$-$CH_3$ is shown. Interestingly, unlike the shorter alkane systems, the second nearest neighbor peak at 5.14 Å in RDF is considerably sharper than the first nearest neighbor peak. We note the presence of a prominent layering structure in $n$ > 4 systems, as illustrated in the respective insets to Figure 2. It is possible that intra-layer molecules might be rotating in synchronization and not contribute to the peak broadening in RDF, while this may not be the case for inter-layer molecules. In contrast, for $n$ < 4 systems molecules are relatively more aligned along the *b*-axis and small enough that the layers are comparable with the inter-layer gap.

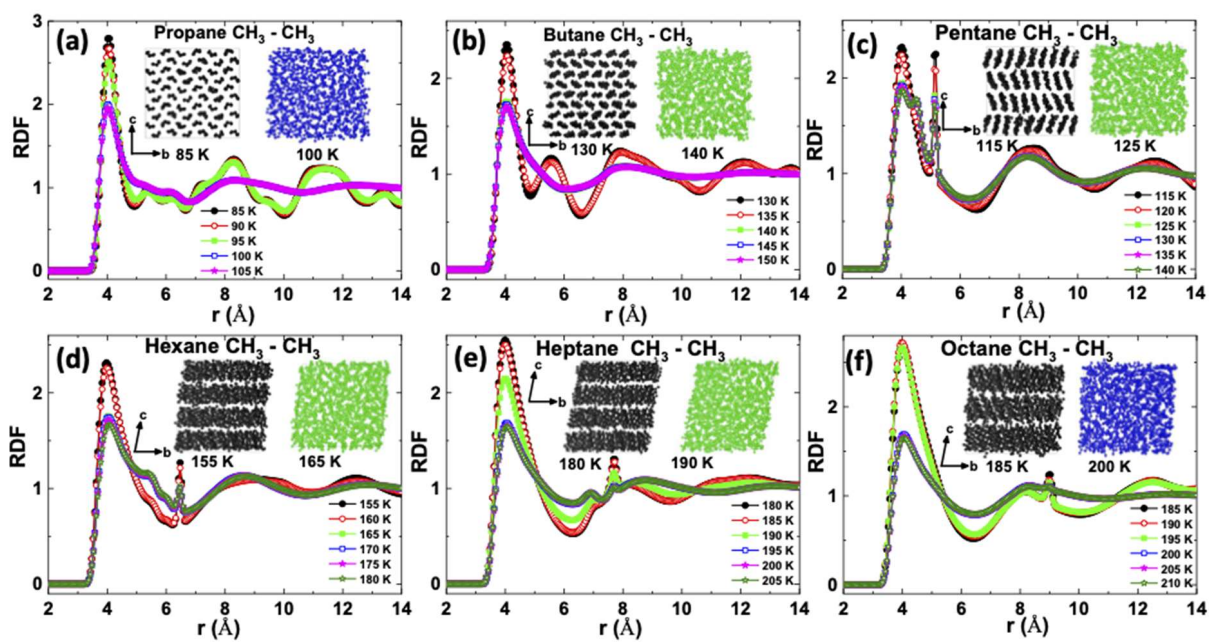

**Figure 2**: Radial distribution function (RDF), for CH$_3$ - CH$_3$ pairs belonging to two different molecules, as a function of temperature for (a) propane, (b) Butane, (c) pentane, (d) hexane, (e) heptane and (f) octane. The insets show the simulation box for a respective n-alkane in crystalline state at low temperature on the left (in black) and in a disordered state at melting point on the right (in blue/green). For all the molecules there is a transition, with increasing temperature, from an ordered crystalline state (indicated by sharp peaks) to a disordered liquid like state (indicated by shallow and broad peaks).

Similar to pentane, a sharp second nearest-neighbor peak in the RDF for hexane (Figure 2(d)), ascribed to the rotational motion of intra-layer molecules, is observed at 6.47 Å. For hexane a shift in second nearest neighbor peak position and together with a reduction in intensity is observed. This can be correlated with increase in the intra-layer molecular spacing with the addition of CH$_3$. Therefore, as a function of increasing $n$ a continuous shift in the second nearest neighbor peak position with concomitant reduction in intensity is observed (Figure 2(e) for $n$ = 7 and 2(f) for $n$ = 8).

To understand the conformational order in alkanes, we calculated the distribution of CH$_3$ - CH$_3$ intramolecular distances. This CH$_3$ - CH$_3$ distance can also be viewed as an effective length of the molecules. In the insets to Figure 3(a) and 3(b) CH$_3$ - CH$_3$ intramolecular distances at (i) in the crystalline state and (ii) at respective melting points is shown, respectively.

A bimodal distribution in the distances is observed for all the alkanes, except for propane ($n$ =3). To probe the odd - even behavior in the bond length distribution below and above the respective melting points we calculated the full width at half maximum (FWHM) of the sharper peaks and distance between the two peaks, as shown in Figure 3(a) and 3(b), respectively. A clear odd - even behavior with no significant variation with temperature, is observed. Interestingly, here heptane ($n = 7$) appears to follow the odd - even trend, unlike its behavior observed in other calculated properties. This may indicate that translational and rotational dynamic behavior is strongly influenced by intermolecular interactions, and hence density as compared to the molecular shape.

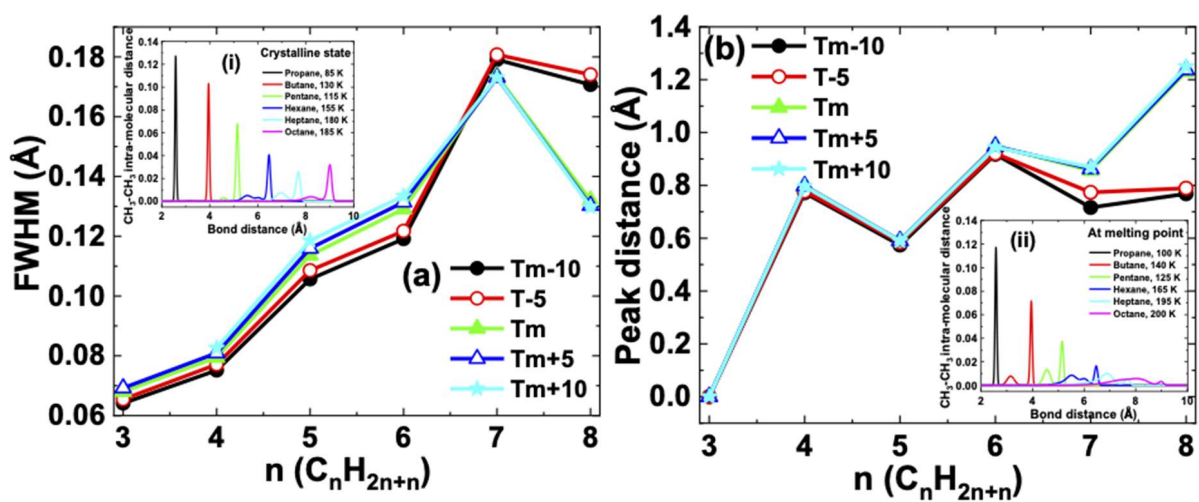

**Figure 3:** (a) Full width half at maximum and (b) distance between the bimodal peaks obtained from bond distance distribution of intramolecular $CH_3$ - $CH_3$ units. The $CH_3$ - $CH_3$ distance is a measure of the effective size of the molecules, with a distribution showing the two preferred sizes. In the insets to Figures (a) and (b) $CH_3$ - $CH_3$ intramolecular distances at (i) in the crystalline state and (ii) at the respective melting points is shown, respectively.

### 3.2. Evolution of dynamics with temperature

### 3.2.1. Translational motion

To understand the translational diffusive behavior of the alkanes, the time dependent mean square displacement (MSD) of the centres of mass of the molecules was calculated. Figure 4(a) depicts the temperature dependent MSD for pentane at both short and long (inset to Figure 4(a)) simulated time scales. The time dependence of the MSD for $n = 3 - 8$ as a function of temperature at shorter and longer (inset) timescales is shown in the Figure S5 of supplemental. At shorter timescales, two different time regimes in MSD are observed. Initially, in the first regime, below 0.3 ps, sharp increase in MSD as a function of time is observed. In this regime MSD can be characterized by MSD $\propto t^2$, corresponding to ballistic motion of the molecules. Here molecules move freely without any interaction with the neighbouring molecules and therefore, no significant influence of temperature on MSD in the probed time scale is observed. In the second regime, above the 1 ns time scale, when the system is below its respective melting point a plateau-like trend with no significant variation in MSD with time is observed. This indicates highly restricted motion under the influence of neighboring molecules, also referred to as a 'caging' region [43]. We observe that at these temperatures, molecules are unable to escape the 'cage' formed by the surrounding molecules even at longer times (up to 10 ns). This implies an absence of the long-range diffusion typical of liquids. At higher temperatures, at and above the respective melting points of the molecules, in the second regime (above 1 ns) not only does the MSD exhibits nearly linear time dependence (MSD $\propto t^1$), but a strong influence of temperature is also seen. The MSD displaying linear behavior as a function of time indicates pure diffusive motion, while sub- or super- diffusive motion would arise from departure from linearity.

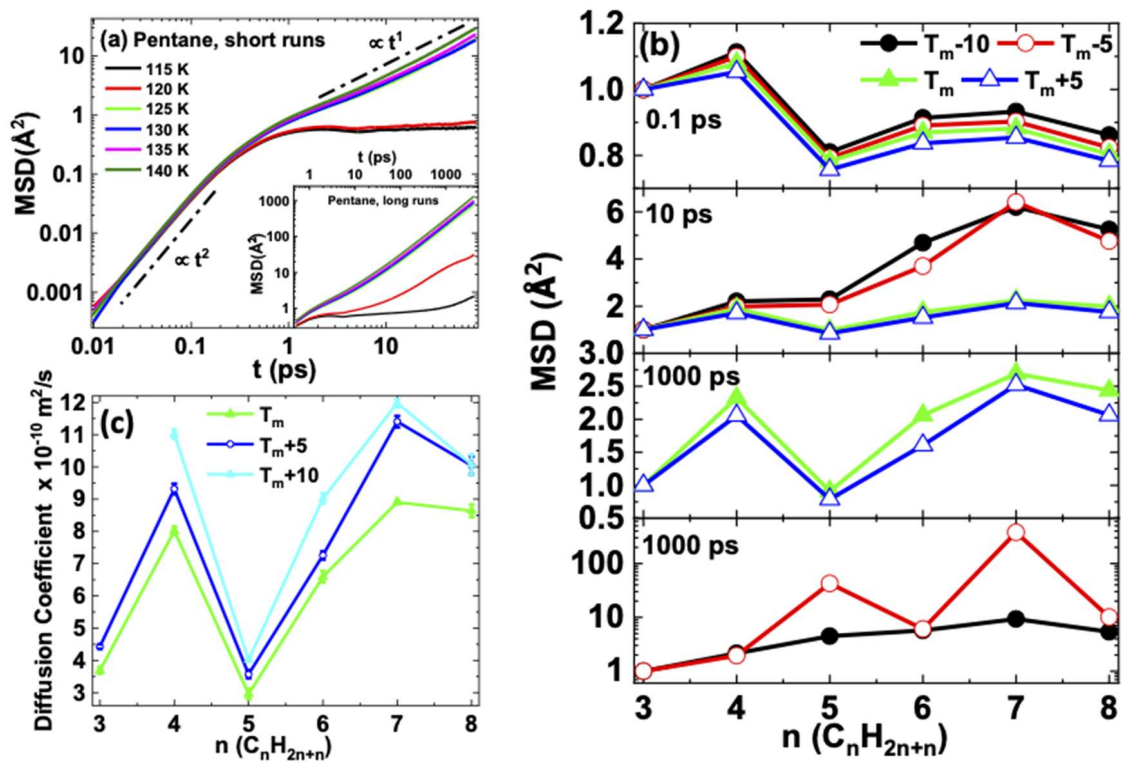

**Figure 4:** (a) Mean square displacement (MSD) variation with temperature for pentane molecule at both short and long (inset to Figure) time scale simulations. Two different time regimes (MSD $\propto t^1$ and MSD $\propto t^2$) are highlighted. (b) Time dependence of the mean square displacement (MSD) at 0.1, 10 and 1000 ps for all alkanes at temperatures below and above the respective melting points. All the data points have been normalized with respect to the MSD value at $T_m - 10$ for propane for each time stamp. All the time regimes show a clear odd - even dynamic behavior, except for $n = 7$. This behavior is in corroboration with the melting point and density behavior depicted in Figure 1. The lines between the points serve as guide to the eye. (c) Diffusion coefficient (D) as a function of $n$ representing long range diffusive motion. An odd - even behavior in D values as a function of $n$ is clearly visible even at temperatures above the melting point, $T_m$. The diffusion coefficient values are obtained by fitting the MSD as a function of time using Einstein equation. The lines between the points serve as guide to the eye.

Figure 4(b) summarizes the time dependence of the overall MSD for all the alkanes at temperatures below and above the respective melting points. For this, the MSD values at 0.1, 10 and 1000 ps are plotted. For clarity all the data points have been normalized with respect to propane MSD value at $T_m$ - 10 for each time stamp. At all the time regimes, a clear odd – even dynamic behavior is observed, again except for $n$ = 7. To further quantify the diffusive behavior, the MSD as a function of time (at longer timescales) was fitted using the Einstein equation, given by equation 2

$$D = \frac{MSD}{2n_d} \qquad (2)$$

where $D$ is the diffusion coefficient and $n_d$ (= 3) represents the dimension. The diffusion coefficient as a function of $n$ is shown in Figure 4(c). As a function of $n$, $D$ exhibits an odd - even behavior both at and above $T_m$, for $T_m$ + 5 and $T_m$ + 10.

3.2.2. Rotational motion

The temperature variation of orientational dynamics can be investigated by studying the evolution of a unit vector $\boldsymbol{u}$ (shown in the inset to Figure 5(a)) along the position vector of a pseudo-atom comprising the molecule with respect to its centre of mass. The orientational correlation functions (OCF) are calculated using equation 3.

$$C_1(t) = \langle \boldsymbol{u}(t + t_0) \cdot \boldsymbol{u}(t) \rangle \qquad (3)$$

A representative figure depicting the typical OCF behavior with time for pentane at several temperatures is shown in Figure 5(a) for the short and longer run simulations (inset to Figure 5(a)). The time dependent OCF behavior is shown for all the alkanes in Figure S6 of supplemental.

Below their respective melting points, all alkanes show non-zero residual correlations at longer times, indicating a severely constrained rotation. These residual correlations suggest that the corresponding molecules are unable to span the entire orientational space and are restricted to undergo libration-like reorientational motion, a characteristic of crystalline structure. This behavior is in agreement with the high degree of anisotropy observed in the ODF at low temperatures (Figure S2 of supplemental). Above the respective melting points, most of the correlations decay to zero within 100 ps. This suggests that at these temperatures the molecules span the entire orientational space via rotational motion, a behavior akin to liquids. It is important to note that for pentane and heptane at 125 K and 190 K respectively, we observe an OCF trend of an intermediate state. This behavior is consistent with the RDF and MSD results described earlier.

Further, the shape of the OCF facilitates identification of two time regimes, one below ~ 1 ps, where the decay of OCF has a relatively steep slope, and another above ~ 1 ps where the slope has flattened. Such distinct regimes in the OCF have been identified earlier with a 'free-rotor' like motion (first, sub-picosecond regime) and a slower rotational motion where inter-molecular interactions start influencing the rotations (second, longer time regime) [44].

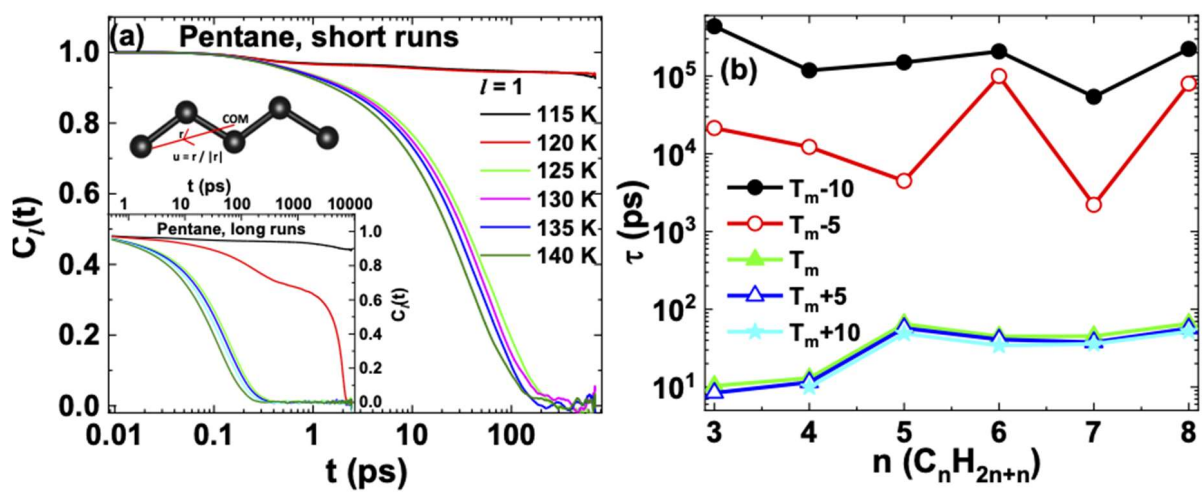

**Figure 5:** (a) The time dependence of orientational correlation function (OCF) at various temperatures for pentane for short simulations and the inset to the figure shows the OCF for longer simulations. Two different time regimes are observed in OCF, below ~ 1 ps corresponding to a 'free-rotor' like motion and above ~ 1 ps describing a slower rotational motion where inter-molecular interactions start influencing the rotations. A unit vector along the position vector of a pseudo-atom comprising the molecule with respect to its centre of mass (COM) is also shown. (b) Variation of the time constant $\tau$ as a function of n in $C_nH_{2n+2}$ above and below melting temperature ($T_m$). Lines joining the points serve as guide to an eye. The time scales ($\tau$) are obtained by fitting the OCF in the regime above ~ 1 ps with an exponential function. The odd – even behavior is most pronounced at temperatures below $T_m$, particularly at $T_m$-5.

We obtained the correlation time scales ($\tau$) by fitting the OCF in the regime above ~ 1 ps with an exponential function. The corresponding variation of $\tau$ as a function of n in $C_nH_{2n+2}$ above and below melting temperature ($T_m$) is depicted in Figure 5(b). Interestingly, the time constant $\tau$ displays odd-even behavior which is distinct at temperatures below $T_m$, particularly at $T_m$-5. At temperatures $T_m$ and above the odd - even behavior appears less pronounced. This behavior is also consistent with the diffusion coefficient obtained from MSD.

## 4.0 Discussion

In a study reported by Gautam et al. [45] fully atomistic (explicit hydrogen model) simulations were performed for propane. When comparing the MSD between two simulation models, not much difference has been observed, implying that flexibility in C - H bonds and H - C - H angles may not have a significant influence on the odd-even behavior.

We compare the melting points of the alkanes obtained from our simulations with experimental data by Boese et al. [5] and Yang et al. [37]. Overall, good agreement between the experimental and simulation results is obtained. Both the melting point and densities follow similar trends. The melting points of even numbered alkanes are higher than the odd ones while, again as expected, the respective densities for even numbered alkanes are lower in comparison to odd alkanes. The translational behavior shown in Figure 4 is in good agreement with the quasi-elastic neutron scattering experimental study reported by Yang et al. [37]. The odd even effect in translational motion obtained from the MSD (Figure 4) is retained at higher temperatures at and above ($T_m$). Also noteworthy is the existence of an odd-even alternation of MSD in the ballistic regime (*t*=0.1 ps). In this regime, the molecules are under free motion and are not affected by the surrounding molecules. An existence of an odd-even effect even for this ballistic motion implies existence of this effect even independent of the nearest neighbor influences and hence the origin of the effect may be the molecular shape itself as suggested by [9]. Similar to this, the existence of odd - even alternation in the molecular size attributes shown in Figure 3 also suggests the importance of molecular shape/conformation in the origin of odd – even effect. This is consistent with [5, 9, 10] that suggest the molecular geometry - trapezoid vs parallelogram as the origin of this effect.

For the rotational dynamics obtained from the OCF (Figure 5) the odd even behavior is only found at temperatures below $T_m$. Therefore, the simulations exhibit the odd-even effect in translational motion even in liquid state, while not in the rotational motion. This implies that odd-even behavior is more prominent at longer time scales.

In further comparison of the structural parameters such as the RDF and ODF, no signature of structural ordering (long or short ranged) is observed above $T_m$. However, the presence of odd-even behavior in translational properties even above $T_m$ indicates that conformational order may not be the only driving force. The intermolecular interactions also contribute to the effect via density. To conclude, both the molecular conformation / geometry, and the intermolecular interactions are important and their interplay gives rise to the odd - even effect.

## 5.0 Conclusions

To obtain a molecular level understanding of the origin of odd-even alteration in the properties of *n*-alkanes, we carried out MD simulations of alkanes with a united atom model. We investigated the structure and dynamics of both positional and orientational order in six *n*-alkanes with $n$ = 3 (propane) to 8 (octane). For each alkane, MD simulations were carried out at several temperatures below and above the melting point, thereby detailing the evolution of the properties and hence their odd-even alternation in temperature across the melting point. Detailed calculations of several properties from the same simulation data helps us see correlations between the odd-even alterations in different properties and thus helps understand the origin of the odd-even effect. It can be concluded from the results reported here that both the molecular shape and density of alkanes play a significant role in the origin of the odd-even effect.

## 6. Author Contributions

Conceptualization, methodology, validation, formal analysis, investigation, writing original draft, project Administration, I. D., M. C. B. and S. G.; Resources, data curation, visualization, I. D. and M. C. B.; Software and coding, S. G.; Review and editing, I. D., M. C. B., S. G., L. P. and J. C. S.

## 7. Acknowledgements

I. D. acknowledge KIFÜ for awarding the access to resource based in Hungary at Budapest. M.C.B acknowledge the independent research fund Denmark for funding received by the international postdoctoral grant under case number: 8028-00023B.

## 8. References

[1] A. Baeyer, Ber. Dtsch. Chem. Ges. 10, 1286–1288 (1877).

[2] E. Badea, G. D. Gatta, D. D' QAngelo, B. Brunetti, Z. R. Re ̌cko ́va, J. Chem. Thermodyn. 38, 1546–1552 (2006).

[3] K. Morishige, T. Kato, J. Chem. Phys. 111, 7095 (1999).

[4] M. K. Mishra, S. Varughese, U. Ramamurty, G. R. Desiraju, J. Am. Chem. Soc. 135, 8121–8124 (2013).

[5] R. Boese, H. Weiss, D. Blaser, Angew. Chem. Int. Ed. 38, 988–992 (1999); Angew. Chem. 111, 1042–1045 (1999).

[6] J. Dupont, J. Braz. Chem. Soc. 15, 341-350 (2004).

[7] A. I. Kitaigorodski, Molekülkristalle, Akademie Verlag Berlin, 380 1979.

[8] F. L. Breusch, Fortsch. Chem. Forsch. 19, 119-184 (1969).

[9] V. R. Thalladi, R. Boese, and H. C. Weiss, Angew. Chem. Int. Ed. 39 918-922, (2000).

[10] Venkat R. Thalladi, Markus Nüsse, and, Roland Boese. J. Am. Chem. Soc. 122, 9227-9236 (2000).

[11] M. T. Mcdermott, J.-B. D. Green, M. D. Porter, Langmuir 13, 2504–2510 (1997).

[12] D. Leythaeuser, R. G. Schaefer, A. Yukler, Am. Assoc. Pet. Geol. Bull. 66, 408–429 (1982).

[13] R. Krishna, J. M. Van Baten, J. Phys. Chem. B 109, 6386–6396391 (2005).

[14] T. L. M. Maesen, E. Beerdsen, S. Calero, D. Dubbeldam, B. Smit, J. Catal. 237, 278–290 (2006).

[15] J. M. Simon, S. Kjelstrup, D. Bedeaux, B. Hafskjold, J. Phys. Chem. B 108, 7186–7195 (2004).

[16] F. Goodsaid-Zalduondo, D. M. Engelman, Biophys. J. 35, 587–594 (1981).

[17] N. G. Almarza, E. Enciso, F. J. Bermejo, J. Chem. Phys. 96, 4625 (1992).

[18] J. A. Pradeilles, S. Zhong, M. Baglyas, G. Tarczay, C. P. Butts, E. L. Myers, V. K. Aggarwal, Nat. Chem. 12, 475–480 (2020).

[19] L. L. Thomas, T. J. Christakis, W. L. Jorgensen, J. Phys. Chem. B 110, 21198–21204 (2006).

[20] K. Yang, M. Tyagi, J. S. Moore, Y. Zhang, J. Am. Chem. Soc. 136, 1268–1271 (2014).

[21] K. Yang, Z. Cai, M. Tyagi, M. Feygenson, J. C. Neuefeind, J. S. Moore, Y. Zhang, Chem. Mater. 28, 3227–3233 (2016).

[22] V.R. Thalladi, R. Boese, H.C. Weiss, J. Am. Chem. Soc. 122, 6, 1186-1190 (2000).

[23] L. Dall'Acqua, G. Della Gatta, B. Nowicka, P. Ferloni, J. Chem. Thermodyn. 34, 1–12 (2002).

[24] M.G. Broadhurst, J. Res. Natl. Bur. Stand., Sect. A 66, 241–249 (1962).

[25] P. N. Nelson, H. A. Ellis, Dalton Trans., 41, 2632-2638 (2012).

[26] H. Zhang, C. Xie, Z. Liu, J. Gong, Y. Bao, M. Zhang, H. Hao, B. Hou, Q. -x. Yin, Ind. Eng. Chem. Res. 52, 18458– 18465 (2013).

[27] M. I. Capar, E. Cebe Phys. Rev. E 73, 061711-061718 (2006).

[28] M.A. A. Rocha, C. M. S. S. Neves, M. G. Freire, O. Russina, A. Triolo, J. A. P. Coutinho, L. M. N. B. F. Santos, J. Phys. Chem. B 117, 10889-10897 (2013).

[29] M. A. R. Martins, C. M. S. S. Neves, K. A. Kurnia, P. J. Carvalho, M. A. A. Rocha, L. M. N. B. F. Santos, S. P. Pinho, M. G. Freire, Fluid Phase Equilib. 407, 188-196 (2016).

[30] J. Leys, M. Wübbenhorst, C. P. Menon, R. Rajesh, J. Thoen, C. Glorieux, P. Nockemann, B. Thijs, K. Binnemans, S. Longuemart, J. Chem. Phys. 128, 064509-064506 (2008).

[31] W. Zheng, A. Mohammed, L. G. Hines, Jr. D. Xiao, O. J. Martinez, R. A. Bartsch, S. L. Simon, O. Russina, A. Triolo, E. L. Quitevis, J. Phys. Chem. B 115, 6572-6584 (2011).

[32] M. A. A. Rocha, J. A. P. Coutinho, L. M. N. B. F. Santos, J. Chem. Phys. 139, 104502-104507 (2013).

[33] M. Vilas, M. A. A. Rocha, A. M. Fernandes, E. Tojo, L. M. N. B. F. Santos, Phys. Chem. Chem. Phys. 17, 2560-2572 (2015).

[34] H. F. D. Almeida, M. G.Freire, A. M. Fernandes, J. A. Lopes-da-Silva, P. Morgado, K. Shimizu, E. J. M. Filipe, J. N. Canongia Lopes, L. M. N. B. F. Santos, J. A. P. Coutinho, Langmuir 30, 6408-6418 (2014).

[35] M. A. A. Rocha, C. F. R. A. C. Lima, L. R. Gomes, B. Schröder, J. A. P. Coutinho, I. M. Marrucho, J. M. S. S. Esperanca, L.P. N. Rebelo, K. Shimizu, J. N. Canongia Lopes, L. M. N. B. F. Santos, J. Phys. Chem. B 115, 10919-10926 (2011).

[36] C. S. Consorti, P. A. Z. Suarez, R. F. de Souza, R. A. Burrow, D. H. Farrar, A. J. Lough, W. Loh, L. H. M. da Silva, J. Dupont, J. Phys. Chem. B 109, 4341-4349 (2005).

[37] K. Yang, Z. Cai, A. Jaiswal, M. Tyagi, J. S. Moore, and Y. Zhang, Angew. Chem., Int. Ed. 55, 14090–14095 (2016).

[38] M. G. Martin, J. I. Siepmann, J. Phys. Chem. B, 102, 2569–2577 (1998).

[39] I. T. Todorov, W. Smith, K. Trachenko, M. T. Dove, J. Mater. Chem. 16, 1911-1918 (2006).

[40] W. Smith and I. T. Todorov, Molecular Simulation 62, 935-943 (2002).

[41] K. Momma and F. Izumi, J. Appl. Crystallogr., 44, 1272-1276 (2011).

[42] W. Humphrey, A. Dalke, K. Schulten, J. Molec. Graphics, 14,133-38 (1996).

[43] J. Geske, B. Drossel, and M. Vogel, AIP Adv. 6, 035131 1-10 (2016).

[44] I. Dhiman, D. Bhowmik, U. Shrestha, D. Cole and S. Gautam, Chem. Eng. Sci., 180, 33-41 (2018).

[45] S. Gautam, A. Kolesnikov, G. Rother, S. Dai, Z.-A. Qiao and D. Cole, J. Phys. Chem. A, 122, 6736-6745 (2018).

# Odd-even effect in n-alkane systems: A molecular dynamics study

I. Dhiman[a], Marcella C. Berg[b, c], Loukas Petridis[c], Jeremy C Smith[c], S. Gautam[d]

[a]Centre for Energy Research, Budapest, 1121 Hungary

[b]Juelich Centre for Neutron Science (JCNS) at Heinz Maier-Leibnitz Zentrum (MLZ), Forschungszentrum Juelich GmbH, Garching, Germany

[c]Oak Ridge National Laboratory, Center for Molecular Biophysics, Oak Ridge, TN, 37831, USA

[d]School of Earth Sciences, The Ohio State University, 275 Mendenhall Laboratory, 125 S Oval Drive, Columbus, OH, USA

## Supplementary Material

| Molecule | Unit cell dimensions | Unit cell multiplicity |
|---|---|---|
| Propane | a = 4.14800 b = 12.61200 c = 6.97700; α = 90.0000 β = 91.2800 γ = 90.0000 | 7 × 3 × 5 |
| Butane | a = 5.70270 b = 5.52470 c = 8.38900; α = 90.0000 β = 115.2200 γ = 90.0000 | 6 × 6 × 4 |
| Pentane | a = 4.13570 b = 9.02500 c = 14.81600; α = 90.0000 β = 90.0000 γ = 90.0000 | 8 × 4 × 2 |
| Hexane | a = 4.13090 b = 4.69630 c = 8.53900; α = 83.4000 β = 87.2650 γ = 75.1720 | 8 × 8 × 4 |
| Heptane | a = 4.11600 b = 4.68600 c = 20.34800; α = 78.1100 β = 81.7900 γ = 74.2500 | 8 × 8 × 2 |
| Octane | a = 4.12300 b = 4.68600 c = 10.97400; α = 85.0600 β = 83.7200 γ = 75.100 | 8 × 8 × 3 |

Table S1: Structural details with unit cell dimensions and the related multiplicity values for n-alkanes.

## Structural Details

The orientation order parameter, s(t), is a calculation characterizing the alignment of a system, suitable for intrinsically anisotropic materials, such as liquid crystals. S is given by the equation below,

$$s(t) = \left\langle \frac{3}{2}\cos\theta(t) - \frac{1}{2} \right\rangle$$

where θ(t) denote the angle made by the molecular axis with respect to the reference direction and ⟨⟩ denotes an average ensemble. For all alkanes, except butane, order parameter value increases from - 0.5 to - 0.2. This implies the change in orientation from anti-parallel / parallel to completely random alignment with respect to the reference direction, indicating that system changes from crystalline to liquid state. For butane, initial crystalline state is in a partially alignment with s = 0.25 and with increasing temperature goes to completely random alignment at s = 0. The calculations are plotted in Figure S1.

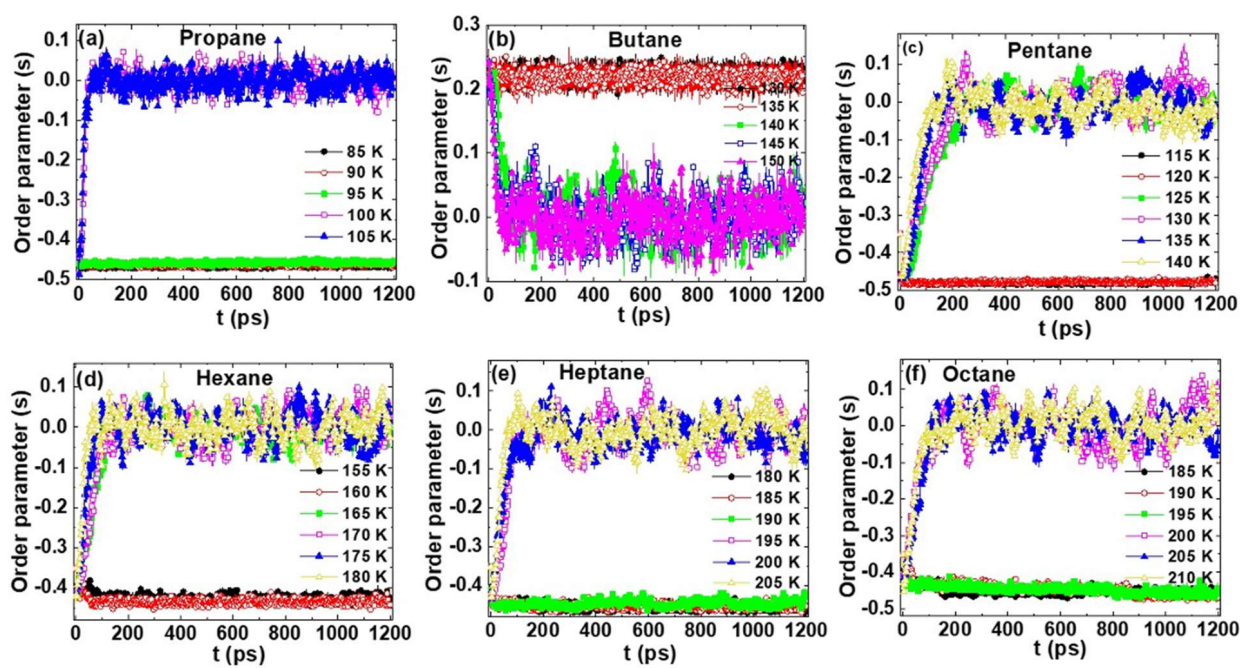

Figure S1: Orientational order parameter (s) calculated for all the alkane as a function of temperature for (a) propane (b) butane (c) pentane (d) hexane (e) heptane and (f) octane molecules.

The orientation distribution function (ODF) is a measure of molecular orientation aligned at an angle (θ) with Cartesian directions (x, y, z). In the crystalline state, molecules depict preferred orientation with respect the Cartesian directions (x, y, z) and therefore the ODF exhibits sharp distinct peaks. In liquid state the orientation is isotropic along all directions, which is represented by shallow and uniform overlapping peaks (for all x, y, and z directions) with a smooth maximum at 90 degrees.

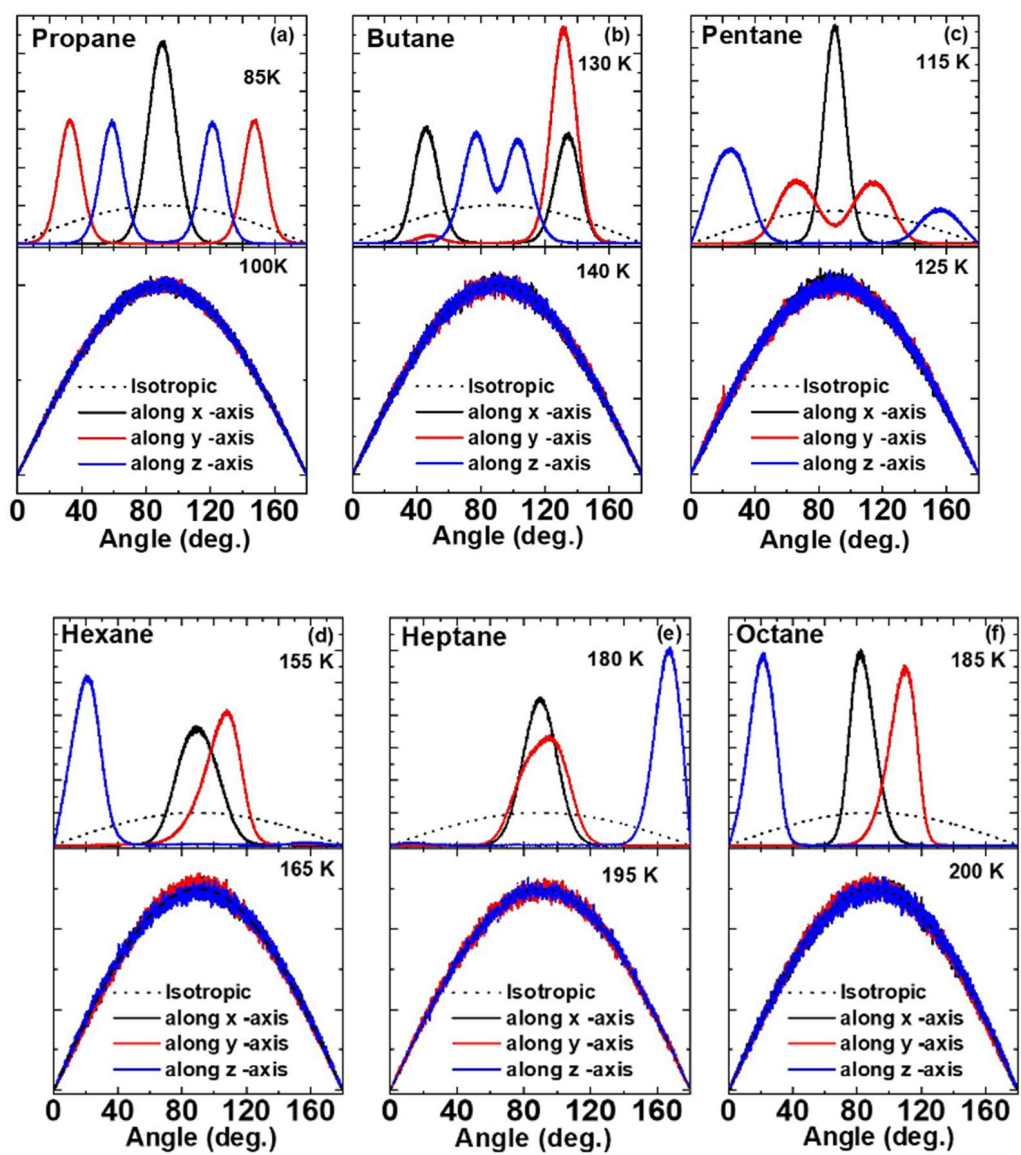

Figure S2: Orientational distribution function for (a) propane (b) butane (c) pentane and (d) hexane (e) heptane and (f) octane molecules along X - (black line), Y - (red line), and Z - (blue line) axis, above and below the respective melting points. The expected orientational distribution for an ideal isotropic behavior for comparison is shown with black dashed lines.

The radial distribution function (RDF) describes the probability distribution of molecules at a given distance, hence providing the local structural information. The temperature dependence of RDF for $CH_3 – CH_2$ and $CH_2 - CH_2$ ($CH_3 - CH_3$ is shown in the main manuscript) pairs for all molecules are shown in Figure S3 and S4, respectively. Similar for all alkanes, we observe a temperature dependent transition from an ordered crystalline state represented by the presence of sharp peaks to a disordered liquid state shown by shallower and broader peaks. The slight broadening of RDF peaks in the ordered state, i. e. below their respective melting points for all the molecules, can be attributed to the presence of rotational disorder, angular and dihedral vibrations in the system.

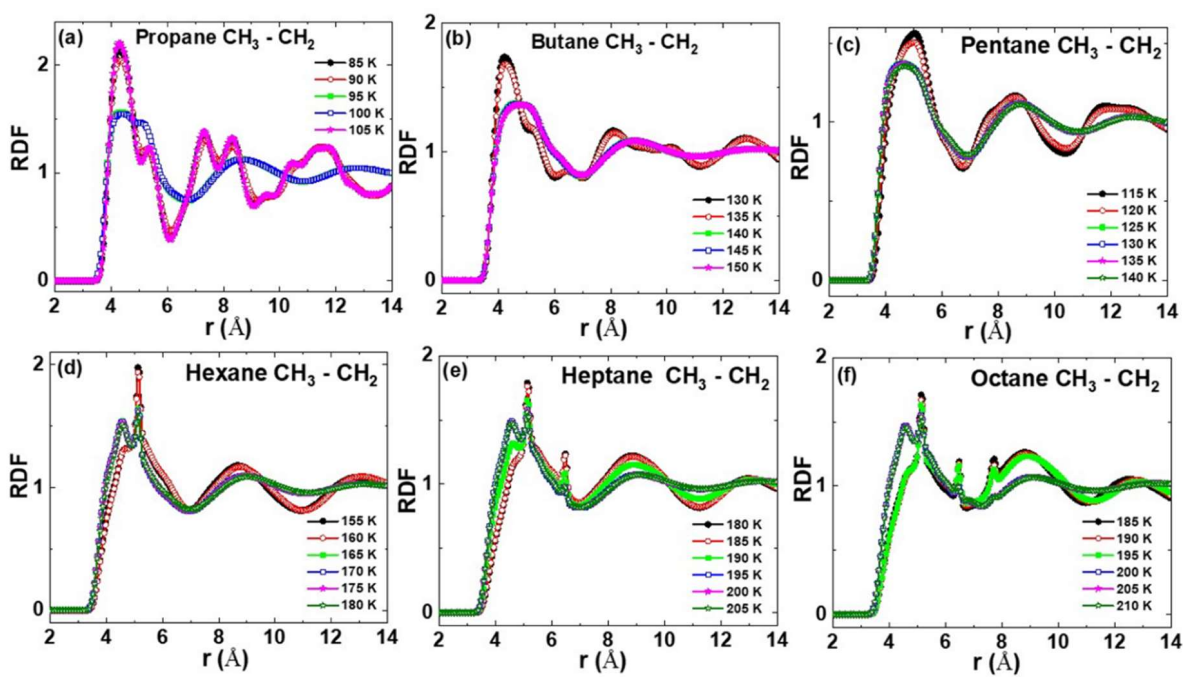

Figure S3: Radial distribution function (RDF) as a function of temperature of $C(H_3)$ - $C(H_2)$ pairs belonging to two different molecules (a) propane (b) butane (c) pentane (d) hexane (e) heptane and (f) octane molecules.

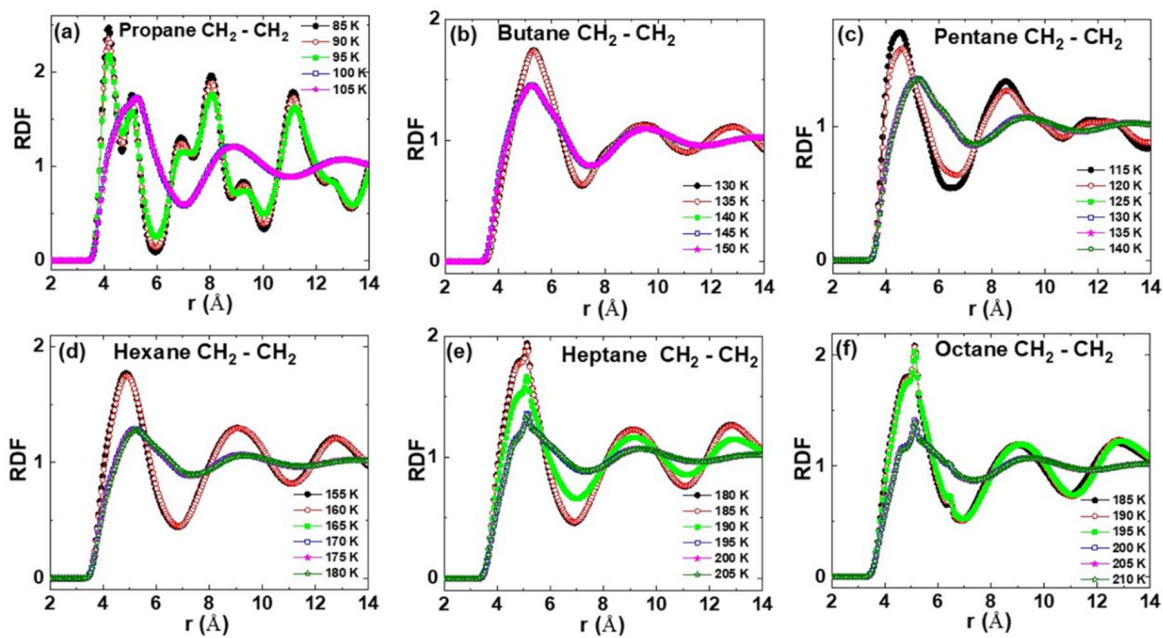

Figure S4: Radial distribution function (RDF) as a function of temperature of $C(H_2)$ - $C(H_2)$ pairs belonging to two different molecules (a) propane (b) butane (c) pentane (d) hexane (e) heptane and (f) octane molecules.

# Dynamical Details

The mean squared displacement (MSD) is a measure of the deviation of a molecule position with respect to its starting position over time. Therefore, the time dependent MSD from the center of mass of the molecules can be used to understand the translational diffusive behavior of alkanes. The overall MSD for n = 3 - 8 as a function of temperature at shorter and longer (inset) timescales is shown in the Figure S5.

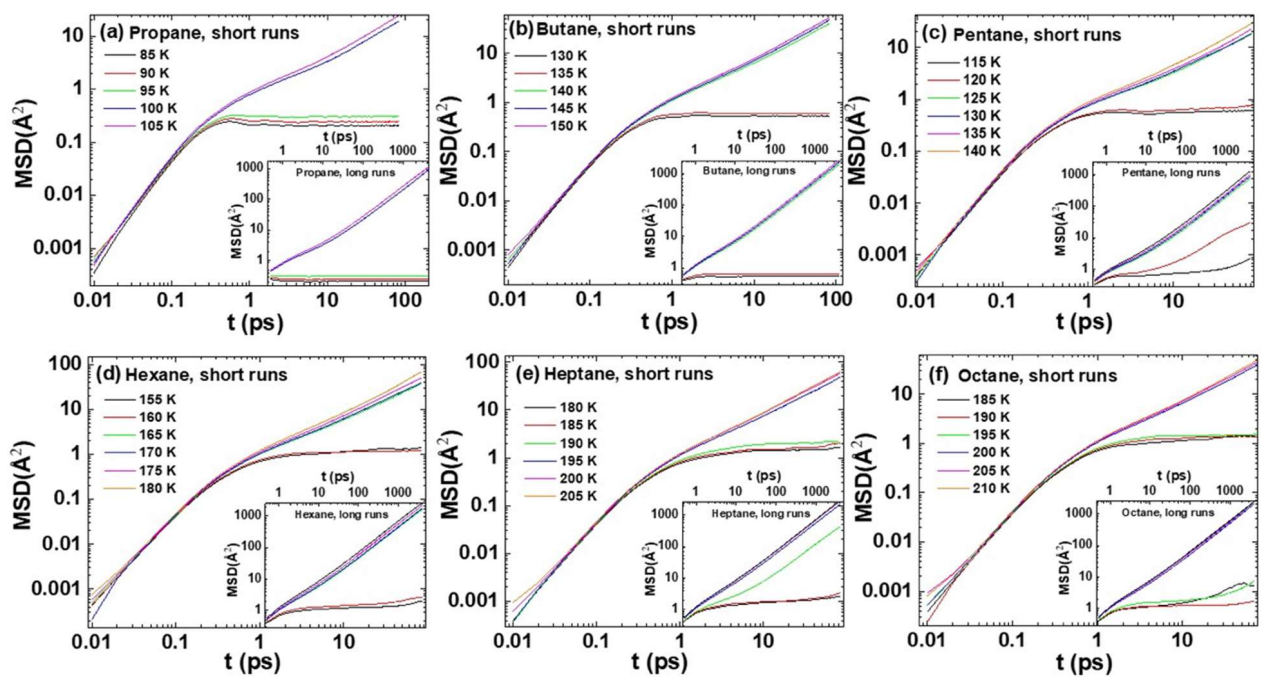

Figure S5: Mean square displacement (MSD) at various temperature for (a) propane (b) butane (c) pentane and (d) hexane (e) heptane and (f) octane molecules. The inset shows the MSD for the respective alkanes at longer timescales (10 ns). Lines between the point serve as guide to the eye.

To investigate rotational motion, the orientational correlation function (OCF) as a function of temperature was used. For all the alkane the time dependent first order ($l = 1$) OCF behavior is shown in Figure S6.

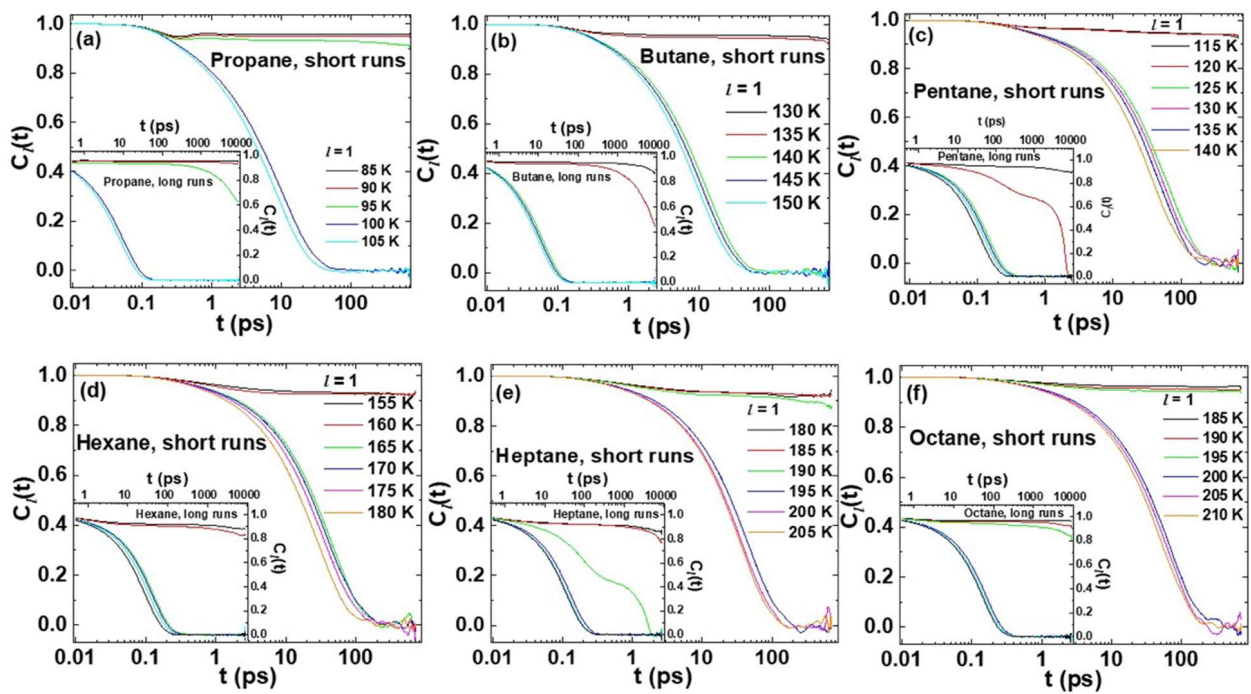

Figure S6: Orientational correlation function (OCF) as a function of temperature for (a) propane (b) butane (c) pentane (d) hexane (e) heptane and (f) octane molecules.